**Hidden memory above the Néel temperature programs antiferromagnetic order in $Cr_2O_3$**

Josiah Keagy, Sumukh Mahesh, Wei-Cheng Liao, and Jing Shi

*Department of Physics & Astronomy, University of California, Riverside, CA 92521, USA*

Control of antiferromagnetic order is central to the development of magnetoelectric and antiferromagnetic spintronic devices. In the prototypical magnetoelectric antiferromagnet $Cr_2O_3$, antiferromagnetic domains are conventionally selected by simultaneous application of electric and magnetic fields during cooling through the Néel temperature ($T_N$) or at a fixed temperature below $T_N$. Here, we show that antiferromagnetic order can instead be programmed by electric-field conditioning performed entirely in the paramagnetic state above $T_N$. By systematically measuring the magnetoelectric coefficient ($\alpha_{zz}$) of bulk $Cr_2O_3$ crystals under different thermal and field histories, we find that applying a sufficiently large electric field above $T_N$ establishes a reversible electrically induced state that persists after the field is removed. Remarkably, subsequent magnetic-field cooling, performed with no applied electric field, produces a single-domain antiferromagnetic state with the opposite Néel-vector orientation from that obtained by conventional magnetoelectric field cooling. The electrically induced state can be reversibly written, erased, and allowed to relax over time at high temperatures, demonstrating a long-lived internal electric bias that governs domain selection during antiferromagnetic ordering. The observed behavior is consistent with a metastable surface-associated polar state that may generate an effective internal electric bias during antiferromagnetic ordering. These results reveal an unexpected coupling between electrically programmed memory and antiferromagnetic domain formation in $Cr_2O_3$.

Corresponding author: Jing Shi

Corresponding author's email: jing.shi@ucr.edu

A central challenge in antiferromagnetic spintronics is the ability to detect and control antiferromagnetic order. Unlike ferromagnets, antiferromagnets possess no net magnetization, making both the detection and manipulation of their order parameter inherently difficult. Magnetoelectric antiferromagnets provide a unique route to overcome this challenge because their antiferromagnetic order is directly coupled to electric and magnetic fields through the linear magnetoelectric effect (MEE) (1-6). $Cr_2O_3$ is the prototypical magnetoelectric antiferromagnet, exhibiting both direct and reverse MEEs (1, 2, 7, 10, 15). These effects have been extensively exploited to detect and control the Néel vector in both bulk crystals and thin films (8-10, 15). In particular, magnetoelectric field cooling (MEFC), in which electric and magnetic fields are simultaneously applied while cooling through the Néel temperature ($T_N$), provides deterministic selection of antiferromagnetic domains and remains the standard approach for programming the Néel vector (1, 3, 9, 11, 12, 15).

Despite decades of study (2-4, 6, 7, 13) the microscopic processes governing antiferromagnetic domain selection in $Cr_2O_3$ remain incompletely understood. In the conventional picture, the Néel vector orientation is determined by the product of the electric and magnetic fields present as antiferromagnetic order develops through $T_N$. This picture implicitly assumes that domain selection is governed only by fields applied during the ordering process itself. Whether information written into the crystal before antiferromagnetic order emerges can subsequently influence domain formation remains largely unexplored.

More broadly, understanding how preexisting electronic, structural, or polar states bias symmetry breaking across a phase transition is an important question that extends beyond magnetoelectric materials. Such hidden memory effects can reveal previously unrecognized order parameters and provide new routes for controlling collective states of matter. In $Cr_2O_3$, surface-sensitive measurements (13) and first-principles calculations have suggested the existence of electrically active surface states and polar surface relaxations (14, 18). Whether such states can influence bulk antiferromagnetic ordering, however, remains unknown.

Here we show that antiferromagnetic domain selection in bulk $Cr_2O_3$ can be programmed by electric-field conditioning performed entirely in the paramagnetic state above $T_N$. Using a reverse magnetoelectric-effect measurement that probes the Néel vector over macroscopic length scales, we find that electric-field annealing above $T_N$ creates a long-lived electrically induced state that persists after removal of the external field and subsequently determines the antiferromagnetic domain selected during magnetic-field cooling. The resulting domain state can be opposite to that obtained through conventional MEFC and can be reversibly written and erased. These observations reveal the existence of a persistent internal electric bias that couples to antiferromagnetic domain formation and suggest a previously unrecognized role of a surface-associated polar state in the control of antiferromagnetic order.

To investigate this effect, we develop a reverse magnetoelectric-effect measurement capable of probing the Néel vector over macroscopic length scales. In our previous work, an anomalous Hall effect-based technique was used to probe nanoscale regions of $Cr_2O_3$ in which a single-domain state is realized. Here we employ a complementary reverse-MEE approach that measures the magnetoelectric response of a macroscopic crystal volume. Fig. 1a illustrates the measurement geometry. An AC magnetic field, $H_\omega$(frequency ~3–7 Hz), is generated using a Kepco power supply that also delivers a DC field $H$ if needed, applied along the crystallographic c-axis ([0001]) of $Cr_2O_3$. The crystal has dimensions of approximately 2.5×2.0×0.25 $mm^3$, with the large surfaces oriented along the (0001) plane. The resulting AC voltage is detected using a pre-amplifier and a lock-in amplifier via two metallic electrodes that cover the entire top and bottom surfaces of the crystal. This voltage is proportional to the oscillating induced polarization $P_\omega$, and hence directly reflects the magnetoelectric coefficient $\alpha_{zz}$. The linear magnetoelectric response is presented in Fig. 1b, where the voltage amplitude is plotted as a function of the AC magnetic field amplitude at 265 K. To further probe the Néel vector control, MEFC is performed by simultaneously applying electric and magnetic fields while cooling the sample from 330 K to ~80 K. The same field magnitudes (1.6 MV/m and 3.4 kOe) are used, but with four different polarity combinations. The resulting temperature dependence of the voltage signals is shown in Fig. 1c.

The four datasets collapse into two mirror-symmetric branches about the horizontal axis, indicating two distinct states corresponding to opposite signs of $\alpha_{zz}$, which are determined by the Néel vector orientation. The voltage reaches its maximum magnitude for both branches, demonstrating that single-domain states are achieved, with the Néel vector uniformly aligned either along the +c or −c direction. These two states are denoted as "$L = +1$" and "$L = -1$," respectively. The table shown alongside Fig. 1c confirms that the Néel vector state is set by the product of the applied electric and magnetic fields, consistent with the linear magnetoelectric coupling. The overall temperature dependence of the voltage follows the expected behavior of $\alpha_{zz}$, as reported in previous studies using alternative techniques (2, 7, 10, 11, 15). In particular, a broad peak is observed, accompanied by a rapid increase in the magnitude just below the Néel temperature ($T_N \approx 307$ K). Once a single-domain state is established, the $\alpha_{zz}$ response remains on one of the two branches during subsequent temperature sweeps below $T_N$, indicating that the single-domain configuration is robust.

Next, we perform electric-field annealing (EA) experiments. Following a procedure analogous to the magnetic field cooling (MFC) sequences, the crystal is first subjected to a static electric field $E_A$ (+1.6 or −1.6 MV/m) for 30 minutes in the paramagnetic state (330 K). The electric field is then removed, the leads are shorted, and the sample is cooled through $T_N$ under a magnetic field of +3.4 or −3.4 kOe. During warming, the leads are disconnected from the short and connected to the measurement electronics for AC voltage detection. As shown in Fig. 2a, two well-defined single-domain branches are obtained, similar to those observed in Fig. 1c, despite the absence of any applied electric field during the cooling process. Notably, the resulting antiferromagnetic single-domain states correspond to the *opposite polarity* of the electric field used during the annealing step at 330 K, as summarized in the table in Fig. 2b. This observation indicates the presence of an effective internal electric bias during MFC. The effective direction of this bias field $E_0$ is opposite to that of the externally applied field during the annealing process. Moreover, the $E_0$ direction can be reversed by reversing the external field direction. The reversible polarity is difficult to reconcile with a simple surface band-bending mechanism (13) and instead points to a more complex origin of the internal field. The persistence of the effect after removal of the external electric field distinguishes it from the conventional magnetoelectric annealing process.

To further test the residual electric-field scenario, we perform the experiment shown in Fig. 3a. We begin with a "+1" single-domain state, prepared via MFC (H=3.4 kOe) following EA with an applied electric field of -1.6 MV/m. The sample is then heated to 330 K, where an annealing field of opposite polarity, +1.6 MV/m, is applied for a controlled duration, to establish a new state. After each annealing period, the electric field is removed, the sample is cooled through $T_N$ by MFC, and the resulting Néel-vector state is determined from the magnetoelectric response voltage. We find that the initially prepared "+1" single-domain state is progressively weakened as the annealing time increases, evolving first into multidomain configurations and eventually settling to the "-1" state after approximately 30 minutes in waiting time. This result indicates that the internal bias responsible for domain selection can be gradually reversed and the new state approaches saturation after ~30 minutes under opposite electric-field annealing.

Fig. 3b shows the relaxation, or wear-off, of the EA-induced state. Starting from either "+1" or "-1" single-domain state prepared by the corresponding EA and MFC procedure as described previously, the sample is heated to 330 K and held for varying durations with neither electric nor magnetic field applied. The sample is then cooled through $T_N$ using MFC, and the resulting Néel-vector state is interrogated by measuring the voltage. We find that the electric-field-induced state responsible for domain selection gradually decays with increasing waiting time above $T_N$. Regardless of the initial domain orientation, prolonged waiting in the absence of any electric field at 330 K leads to a mixed-domain state after MFC, indicating that the electrically induced contribution relaxes over time toward a state characterized by a stable nonzero internal bias in the absence of an applied electric field (3).

The internal electric-field ($E_0$) scenario is further supported by the phase diagram shown in Fig. 4a. The phase boundary is displaced to an applied field of approximately $+0.6$ MV/m, corresponding to an inferred internal bias $E_0 \approx -0.6$ MV/m. Starting from the same initial "-1" state (red region with red circles), the applied electric field during MEFC must first overcome an effective residual field of approximately $E_0$ ~ -0.6 MV/m (green vertical region) before the Néel vector is programmed to the "+1" state (blue region with blue circles). The electric-field dependence for H = 0.5 kOe is shown by the trace above the main panel, and similar threshold behavior is observed at other magnetic fields. This offset of the phase boundary is consistent with the presence of an effective internal electric field opposing the externally applied electric field. Similar offsets have previously been reported in $Cr_2O_3$ thin films (9, 16, 17), although their microscopic origin need not be identical to that observed here.

We note that the transition region between the "-1" and "+1" states is extremely narrow, indicating that only a small product of effective electric and magnetic fields is required to determine the Néel-vector orientation. This behavior is expected because, during MEFC, antiferromagnetic order is established only upon crossing $T_N$, where the anisotropy barrier can be vanishingly small and the linear magnetoelectric coupling can efficiently select the domain state.

Further evidence for $E_0$ is provided by magnetic-field cooling in the absence of an applied electric field. As shown in Fig. S2, starting from either the "+1" or "-1" initial state, only a small magnetic field is required to set the crystal into the opposite single-domain state during MFC. This observation implies that a finite residual electric field persists even when no external electric field is applied. Consequently, the combined action of the residual electric field and the applied magnetic field determines the final Néel-vector orientation through the linear magnetoelectric effect. This interpretation is also consistent with our previous observation of reversible Néel-vector toggling by MFC alone in bulk $Cr_2O_3$ crystals (8).

The observation of a reversible internal field naturally raises the question of its microscopic origin. While charge trapping at interfaces may contribute to the observed behavior, several features—including the long relaxation time even with the leads shorted, reproducible polarity reversal, the persistence of a biased multidomain state after prolonged relaxation (Fig. 3b), and reversible write–erase behavior—suggest that transient charge accumulation alone may not fully account for the observed memory effect. Instead, the observations are more consistent with a metastable structural or dipolar state that can be repeatedly programmed by the applied electric field. Among the possible mechanisms, a surface-associated polar state is particularly attractive because (i) surface relaxations and ionic displacements are predicted for $Cr_2O_3$(0001) (14, 18, 19), (ii) the inferred internal field has the polarity expected for a depolarizing field, (iii) the effect is reversible under opposite electric-field conditioning, and (iv) the surface-associated polar state can generate a local electric bias near the crystal surface. The magnitude of the inferred internal field can be accounted for by sub-percent ionic displacements within a near-surface region. Because antiferromagnetic order is established through domain nucleation and growth near $T_N$, such a local bias may be sufficient to influence the domain that subsequently propagates throughout the crystal. In this picture, the internal bias need not permeate the entire crystal volume; it only needs to influence the initial nucleation of antiferromagnetic domains near $T_N$. Based on the observations, we propose the following physical picture (Fig. 4b). We use the term 'surface-associated polar state' generically to describe a metastable near-surface dipolar configuration, without implying bulk ferroelectric order. At 330 K, the application of a strong electric field during EA may reconfigure or select a metastable surface-associated polar state, which in turn generates an opposing depolarizing field. The induced polarization can persist for a finite duration after the external field is removed. Because $Cr_2O_3$ is not a ferroelectric material, this polarization does not originate from a bulk ferroelectric instability. Instead, it is more plausibly associated with field-induced ionic displacements confined to a near-surface region. If MFC is performed before complete relaxation occurs, the residual depolarizing field combines with the applied magnetic field through the linear magnetoelectric effect and biases antiferromagnetic domain selection as the system cools through $T_N$.

Further evidence for a metastable polar state is provided by the dependence on the annealing-field magnitude measured in two samples, crystals #1 and #2 (Fig. S5). Crystal #1 exhibits a sharp hysteresis loop with well-defined bistable states, while qualitatively similar hysteretic behavior is reproduced in crystal #2, albeit with a broader transition. The remanence, hysteresis, and saturation observed as a function of $E_A$ are characteristic of a bistable polar response and are consistent with the proposed metastable polar-state picture. Although these observations do not exclude charge trapping or other interfacial mechanisms, they place substantially stronger constraints on a simple transient-charge interpretation.

Surface-localized ionic displacements are known to produce ferroelectric-like behavior in otherwise non-ferroelectric oxides, as demonstrated in $SrTiO_3$ (20, 21). Similarly, first-principles calculations predict inward relaxation of the outermost Cr and O layers on $Cr_2O_3$(0001), resulting in a finite surface dipole even in the absence of an applied electric field (14, 18). The long relaxation times observed in Fig. 3b are compatible with a metastable structural or polar configuration, although they do not by themselves distinguish such a state from long-lived electronic or interfacial charge trapping. While the present measurements do not uniquely determine the microscopic surface termination, the observed polarity of the residual field is consistent with a surface dipole that favors the "−1" antiferromagnetic state during MFC. Future structural probes capable of resolving sub-percent surface displacements will be valuable for directly testing the proposed microscopic origin of the internal bias.

In summary, we have observed a memory effect through reverse magnetoelectric measurements that reveals a long-lived internal electric bias capable of influencing antiferromagnetic domain selection during cooling through $T_N$. The inferred internal field is consistent with a metastable near-surface polar state associated with atomic displacements and can be written, erased, reversed, and allowed to relax over time at elevated temperatures. More broadly, our results show that metastable states programmed in the paramagnetic phase can influence symmetry breaking during a later phase transition. This work establishes electric-field conditioning above $T_N$ as a new route for programming antiferromagnetic order and points to a broader paradigm in which hidden memory encoded before a phase transition governs the ordered state that subsequently emerges.

We thank Haoyu Liu, Jia-Mou Cheng, Nick Mancini, Ward Beyermann, and Igor Barsukov for their assistance and helpful discussions. Work at UCR on device fabrication and magnetoelectric coefficient measurements was supported by National Science Foundation under Award Number DMR-2203134, on measurement system setup and materials characterization by the Department of Energy under Award Number DE-FG02-07ER46351, and on data analysis by funds from the UC National Laboratory Fees Research Program of the University of California under Grant Number L25CR9003.

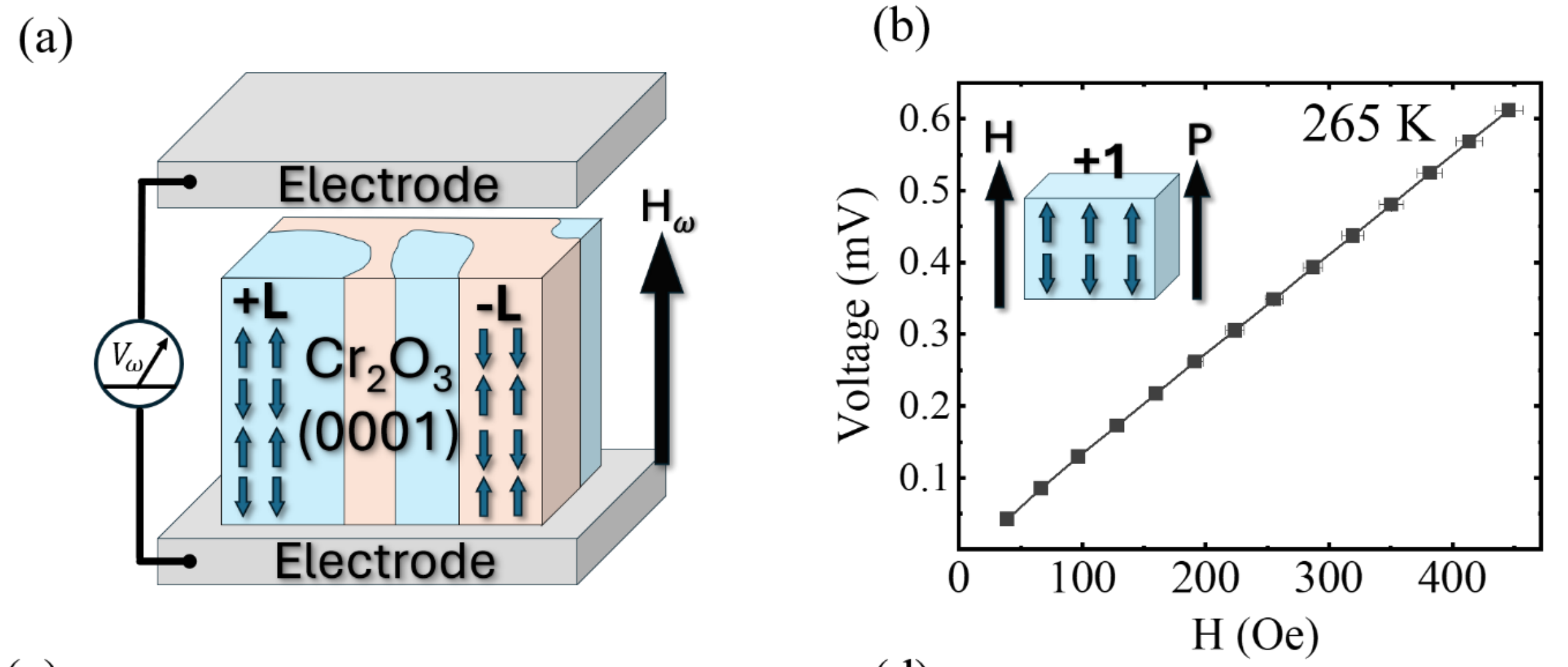

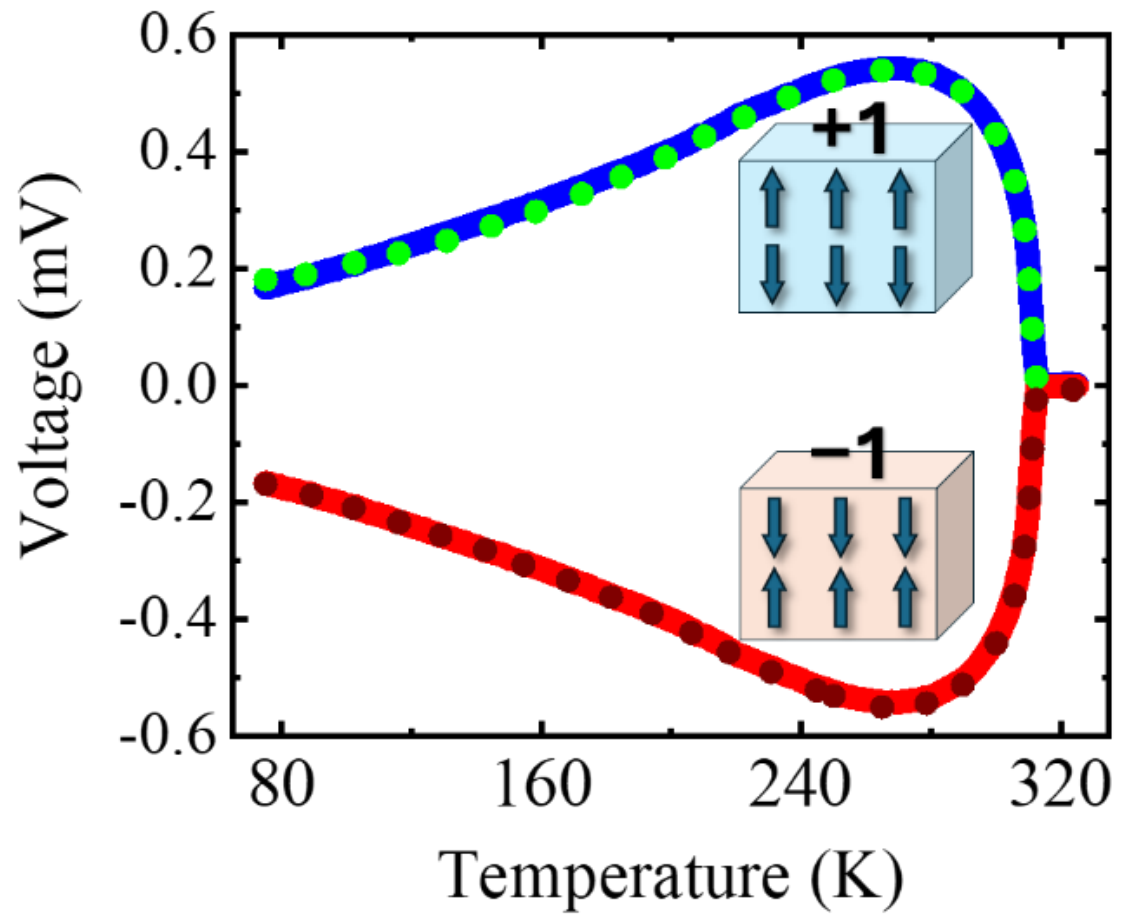

(d)

| | E (MV/m) | H (kOe) | L |
|---|---|---|---|
| | +1.6 | +3.4 | +1 |
| | −1.6 | +3.4 | −1 |
| | −1.6 | -3.4 | +1 |
| | +1.6 | -3.4 | −1 |

**Figure 1**. **Reverse-MEE measurement technique and MEFC.** (a) Schematic of the reverse magnetoelectric-effect measurement on a 2.5 × 2.0 × 0.25 mm³ $Cr_2O_3$ crystal. The apparent gap between the top electrode and the crystal surface is included solely for illustrative purposes to make the antiferromagnetic domains visible; no physical gap is present in the actual device. An AC magnetic field is applied along the easy axis while lock-in measures induced voltage across the crystal. Electric and magnetic fields are applied along the crystallographic c-axis during cooling through $T_N$. Lock-in measurements were taken during warming. (b) AC voltage response as a function of the applied AC magnetic field amplitude (both RMS) at 265 K for single-domain state "+1". The linear dependence confirms the reverse magnetoelectric effect, and the slope is proportional to the magnetoelectric coefficient $\alpha_{zz}$ . (c) Temperature dependence of the reverse magnetoelectric response following MEFC under different combinations of electric and magnetic field polarities. The sign of the signal reflects the orientation of the Néel vector. (d) Summary of the single-domain states obtained under the corresponding MEFC conditions.

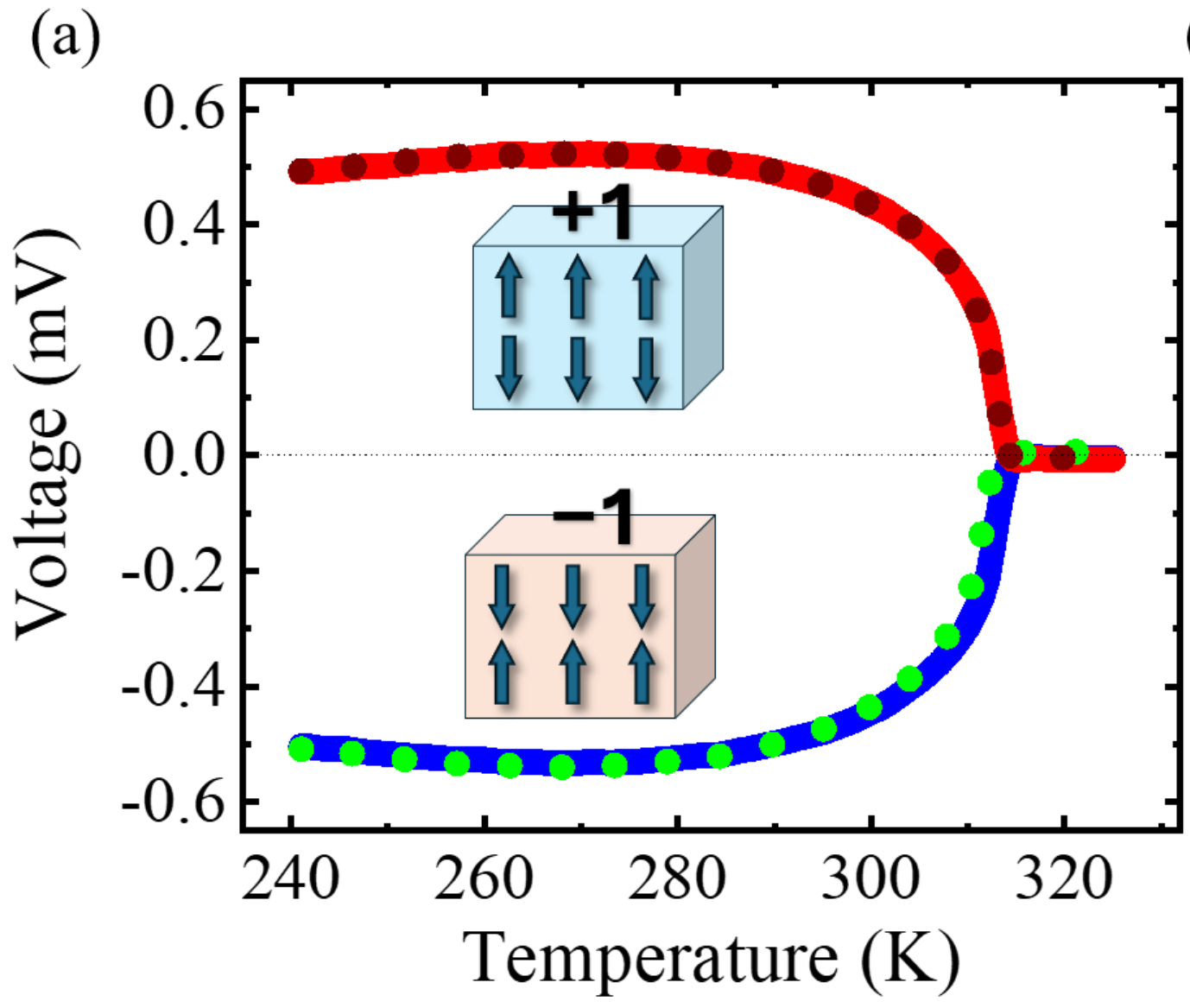

(b)

| | $E_A$ (MV/m) | H (kOe) | L |
|---|---|---|---|
| | +1.6 | +3.4 | −1 |
| | −1.6 | +3.4 | +1 |
| | −1.6 | −3.4 | −1 |
| | +1.6 | −3.4 | +1 |

**Figure 2. Electric-field annealing programs antiferromagnetic order.** (a) Temperature-dependent reverse magnetoelectric response measured during warming after electric-field annealing (EA) at 330 K with $E_A$ = ±1.6 MV/m for 30 min, followed by MFC. Well-defined single-domain states are obtained despite the absence of an applied electric field during cooling, demonstrating that electric-field conditioning in the paramagnetic state determines the subsequent antiferromagnetic domain orientation. (b) Summary of the single-domain states obtained after EA and MFC. The resulting Néel-vector orientation exhibits the opposite polarity to that produced by conventional MEFC shown in Fig. 1d, indicating the presence of an effective internal electric field generated by the prior electric-field annealing.

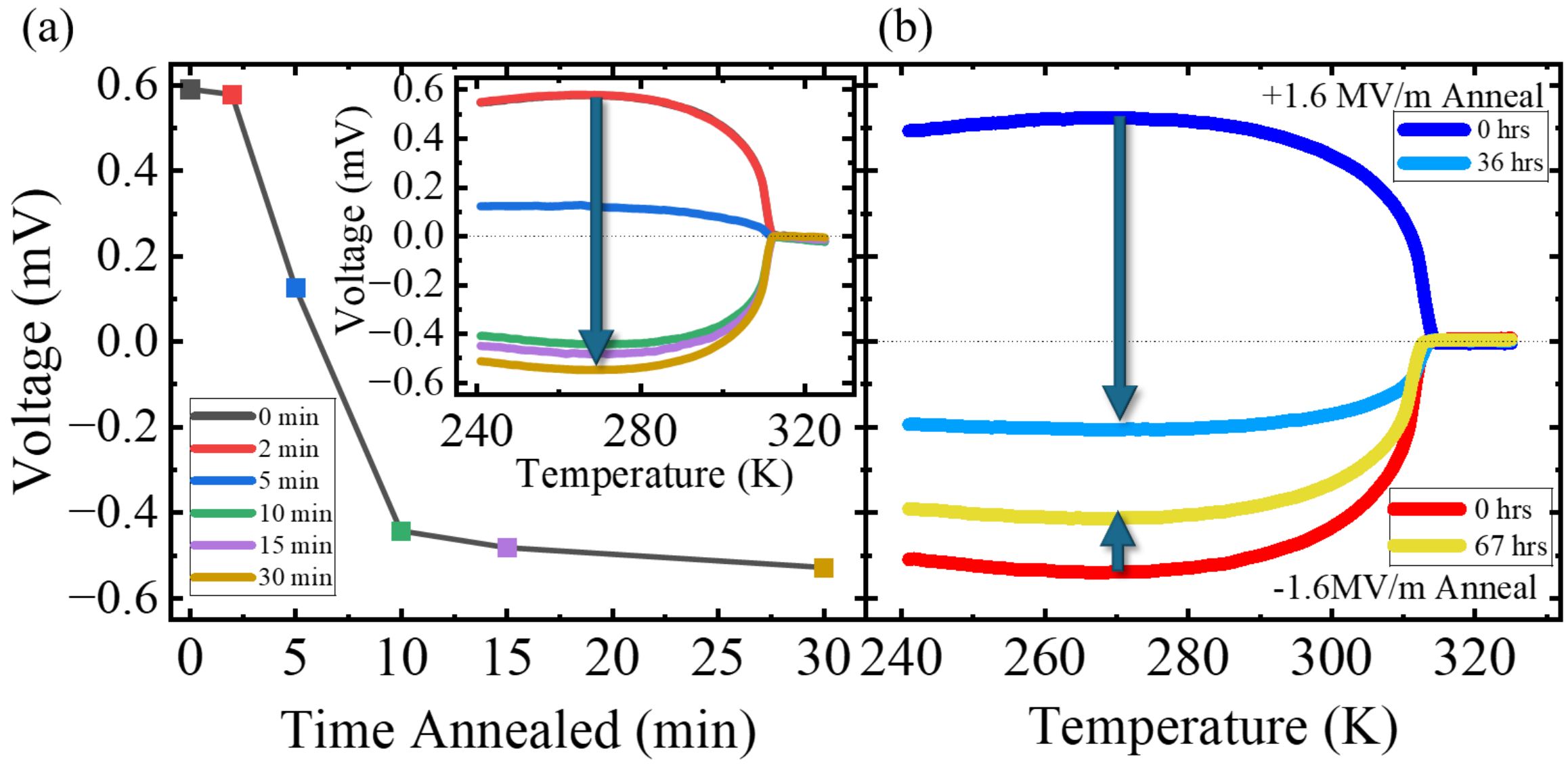

**Figure 3. Writing and erasure dynamics of the electrically induced memory state.** (a) Temperature-dependent reverse magnetoelectric response measured after MFC with H=+3.4 kOe (inset). The initial "+1" single-domain state was prepared by EA at 330 K with $E_A$ = -1.6 MV/m for 30 min. Subsequent annealing with the opposite electric field $E_A$ =+1.6 MV/m for different durations progressively weakens the initial state, driving the crystal through intermediate multidomain configurations before reaching the opposite "-1" single-domain state. The voltage measured at 265 K is plotted as a function of annealing time, showing that the electrically induced state approaches saturation after approximately 30 min. (b) Relaxation of the electrically induced memory state. Following preparation of either the "+1" or "-1" single-domain state by EA and MFC, the crystal was held at 330 K with no applied electric or magnetic field for varying durations before MFC with H=-3.4 kOe. The influence of the prior electric-field conditioning gradually diminishes over the course of several days, and both initial states evolve toward a similar multidomain configuration with a bias toward the "-1" state. The convergence of the two branches indicates the slow relaxation of the electrically induced memory state in the absence of an applied electric field, suggesting the existence of a persistent built-in internal electric bias.

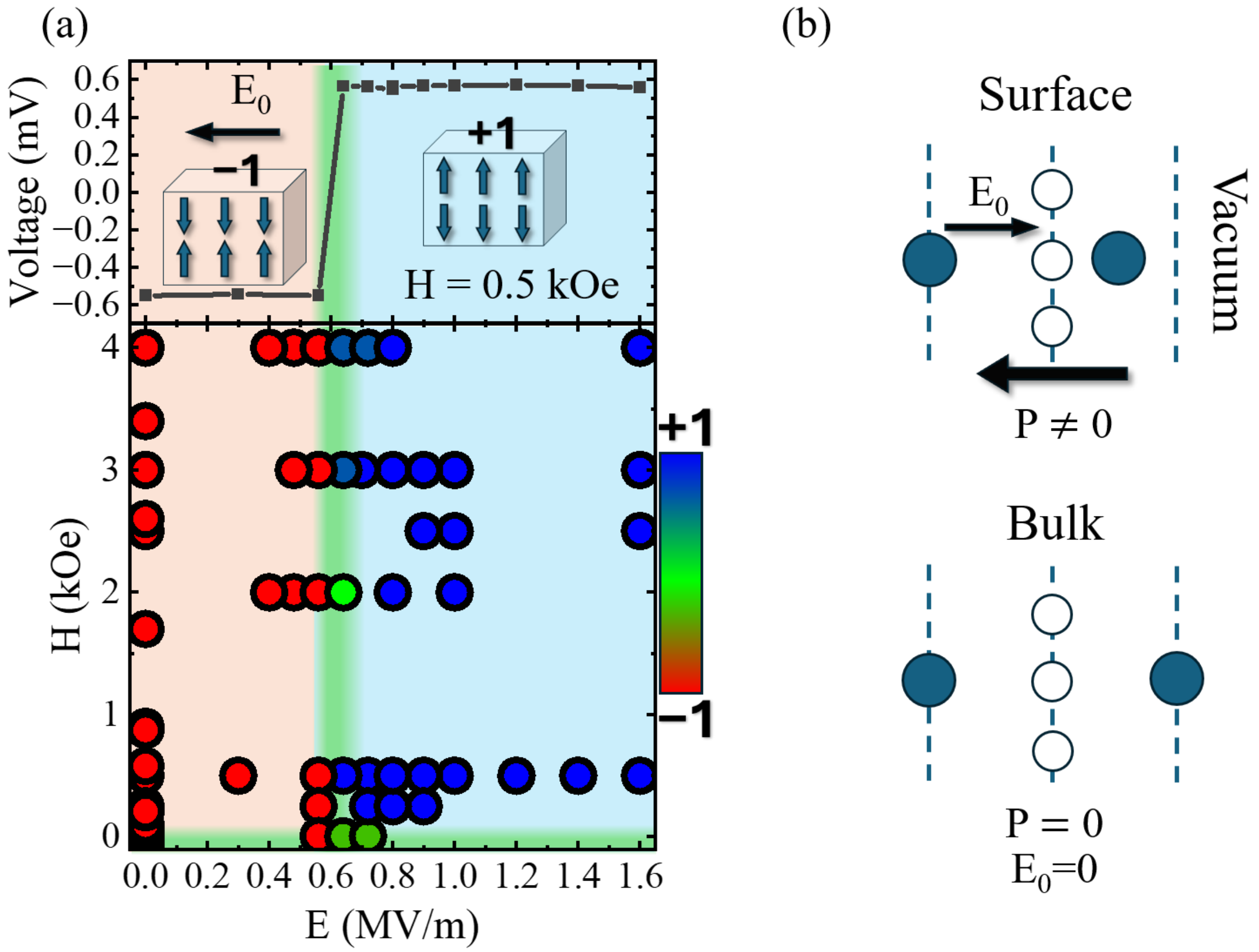

**Figure 4. Evidence for an internal electric field and proposed microscopic origin.** (a) Antiferromagnetic domain states obtained after EA at +1.6 MV/m, followed by MEFC under different combinations of electric and magnetic fields. The voltage measured at 265 K identifies either the "-1" (red circles) or "+1" (blue circles) single-domain state. The narrow vertical transition region (green) indicates a threshold electric field required to overcome an effective internal electric field, $E_0$. The voltage measured after MEFC with H=0.5 kOe is plotted as a function of applied electric field above the main panel. The offset of the phase boundary yields an effective internal field of ~ -0.6 MV/m. (b) Schematic illustration of the proposed microscopic origin of the internal electric field. Surface relaxation of the $Cr_2O_3$ surface leads to inward displacement of the outermost Cr atoms (solid circles), generating a surface-associated polar state extending over multiple atomic layers. Although spontaneous electric polarization is forbidden in the centrosymmetric bulk crystal, inversion symmetry is broken at the surface, allowing a net surface polarization P and the associated depolarizing field $E_0$. The surface polar state can be modified by an applied electric field E.

References

1 S. Shtrikman, D. Treves, Observation of the Magnetoelectric Effect in $Cr_2O_3$ Powders, *Phys. Rev.* **130**, 986-988 (1963).

2 D. N. Astrov, Magnetoelectric Effect in Chromium Oxide, *Sov. Phys. JETP* **13**, 729-733 (1961).

3 T. J. Martin, J. C. Anderson, Magneto-Electric Annealing Effects of $Cr_2O_3$, *Phys. Lett.* **11**, 109-110 (1964).

4 I. E. Dzyaloshinskii, On the Magneto-Electrical Effect in Antiferromagnets, *Sov. Phys. JETP* **10**, 628-629 (1959).

5 P. Borisov, A. Hochstrat, X. Chen, W. Kleemann, C. Binek, Magnetoelectric Switching of Exchange Bias, *PRL* **94**, 117203 (2005).

6 T. J. Martin, J. C. Anderson, Antiferromagnetic Domain Switching in $Cr_2O_3$, *IEEE Transactions on Magnetics* **2**, 446-449 (1966).

7 G. T. Rado, V. J. Folen, Observation of the Magnetically Induced Magnetoelectric Effect and Evidence for Antiferromagnetic Domains, *PRL* **7**, 310-311 (1961).

8 W. C. Liao, H. Liu, W. Tan, J. Keagy, J. M. Chen, J. Shi, Electrical Detection of Single-Domain Néel Vector Reorientation across the Spin-Flop Transition in $Cr_2O_3$ Crystals, *Nano Lett.* **25**, 9794-9800 (2025).

9 T. Kosub, M. Kopte, R. Hühne, P. Appel, B. Shields, P. Maletinsky, R. Hübner, M. O. Liedke, J. Fassbender, O. G. Schmidt, D. Makarov, Purely antiferromagnetic magnetoelectric random access memory, *Nat Commun* **8**, 13985 (2017).

10 H. Wiegelmann, A. G. M. Jansen, P. Wyder, J. P. Rivera, H. Schmid, Magnetoelectric effect of $Cr_2O_3$ in strong static magnetic fields, *Ferroelectrics* **162**, 141-146 (1994).

11 B. B. Krichevtov, V. V. Pavlov, R. V. Pisarev, Nonreciprocal optical effects in antiferromagnetic $Cr_2O_3$ subjected to electric and magnetic fields, *Zh. Eksp. Teor. Fiz.* **94**, 284-295 (1988).

12 T. J. Martin, Antiferromagnetic Domain Switching in $Cr_2O_3$, *Phys. Lett.* **17**, 83-85 (1965).

13 K. Du, X. Xu, C. Won, K. Wang, S. A. Crooker, S. Rangan, R. Bartynski, S. W. Cheong, Topological surface magnetism and Néel vector control in a magnetoelectric antiferromagnet, *npj Quantum Mater.* **8**, 17 (2023).

14 J. A. Cline, A. A. Rigos, T. A. Arias, *Ab initio* study of magnetic structure and chemical reactivity of $Cr_2O_3$ and its (0001) surface, *Journal of Physical Chemistry B* **104**, 6195-6201 (2000).

15 P. Borisov, T. Ashida, T. Nozaki, M. Sahashi, D. Lederman, Magnetoelectric properties of 500-nm $Cr_2O_3$ films, *Phys. Rev. B* **93**, 174415 (2016).

16 K. Ujimoto, H. Sameshima, K. Toyoki, T. Moriyama, K. Nakamura, Y. Kotani, M. Suzuki, I. Iino, N. Kawamura, R. Nakatani, Y. Shiratsuchi, Giant gate modulation of antiferromagnetic spin reversal by the magnetoelectric effect, *NPG Asia Mater* **16**, 20 (2024).

17 H. Sameshima, K. Ujimoto, N. Murayama, K. Toyoki, K. Nakamura, T. Moriyama, R. Nakatani, Y. Shiratsuchi., Highly efficient deterministic voltage control of the perpendicular Néel vector in magnetoelectric antiferromagnet, *Phys. Rev. Applied* **25**, 014017 (2026).

18 C. Rehbein, N. M. Harrison, A. Wander, Structure of the $\alpha$-$Cr_2O_3$ (0001) surface: An *ab initio* total-energy study, *Phys. Rev. B* **54**, 14066 (1996).

19 X. H. Verbeek, A. Urru, N. A. Spaldin, Hidden orders and (anti-)magnetoelectric effects in $Cr_2O_3$ and $\alpha$-$Fe_2O_3$, *Phys. Rev. Research* **5**, L042018 (2023).

20 R. Sachs, Z. Lin, J. Shi, Ferroelectric-like $SrTiO_3$ surface dipoles probed by graphene, *Sci Rep* **4**, 3657 (2014).

21 N. Bickel, G. Schmidt, K. Heinz, K. Müller, Ferroelectric relaxation of the $SrTiO_3$(100) surface, *Phys. Rev. Lett.* **62**, 2009-2011 (1989).

Supplementary Information for

# Hidden memory above the Néel temperature programs antiferromagnetic order in $Cr_2O_3$

Josiah Keagy *et al.*

*Corresponding author. Email: jing.shi@ucr.edu

**This PDF file includes:**

Figs. S1 to S5

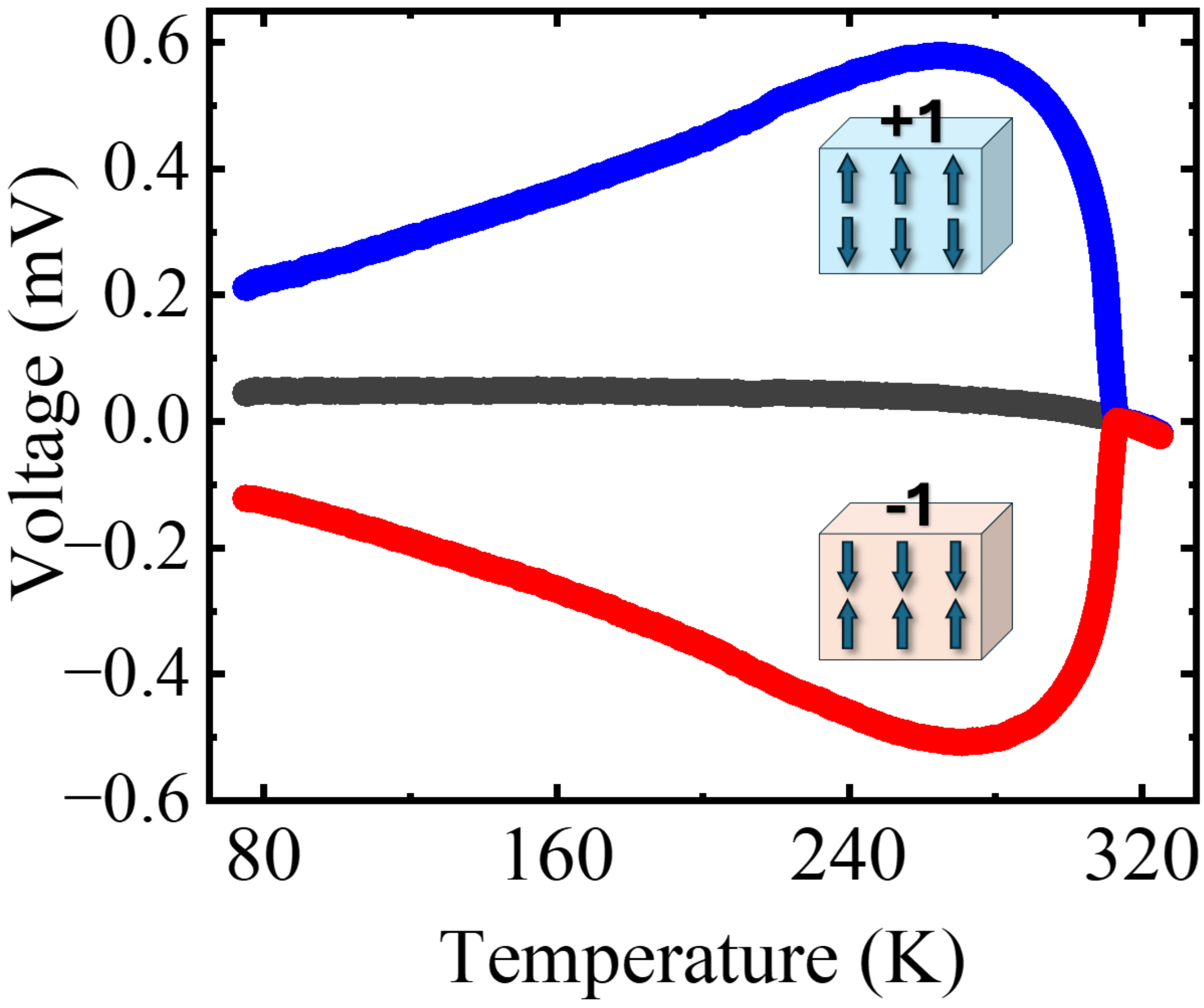

**Fig. S1. Background subtraction of the reverse magnetoelectric signal.** The Néel vector was initialized into the (L= +1) and (L= -1) single-domain states using magnetoelectric field cooling (MEFC), and the corresponding temperature dependence of the reverse magnetoelectric response was measured (blue and red curves, respectively). The background signal (gray curve) was obtained by averaging the responses of the two opposite single-domain states. The absence of the characteristic temperature dependence of the magnetoelectric coefficient in the averaged signal confirms that the background is non-magnetoelectric in origin, likely due to a small temperature-dependent inductive response of the circuit; therefore, it can be reliably separated from the Néel-vector-dependent response.

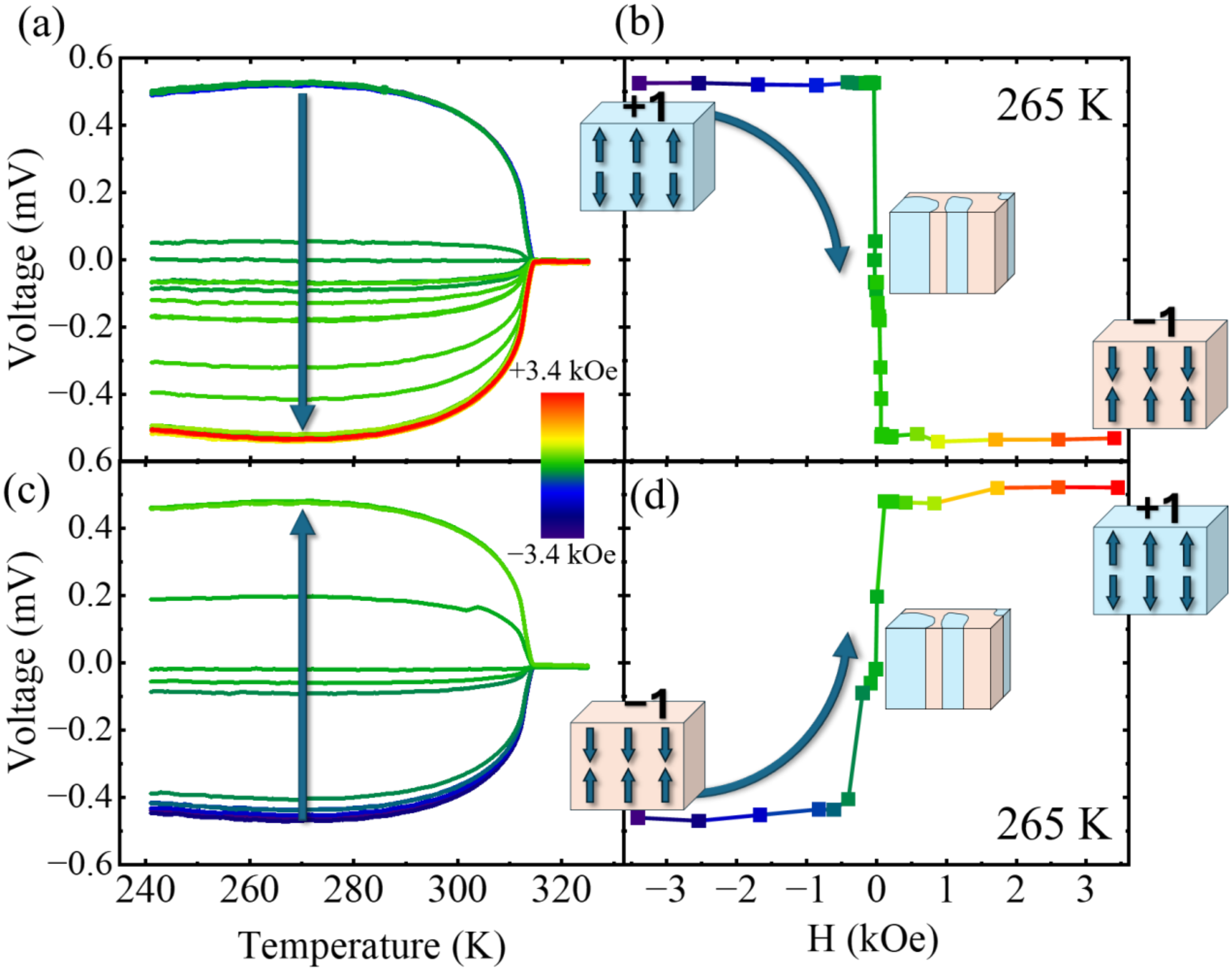

**Fig. S2. Néel-vector toggling by magnetic-field cooling (MFC) following electric-field annealing (EA).** (a) Temperature-dependent reverse magnetoelectric response measured during warming after EA at 330 K with $E_A$ = +1.6 MV/m for 30 min, followed by MFC under different magnetic fields. MFC with -3.4 kOe prepares the initial "+1" single-domain state, followed by MFC sequence with -2.55, -1.7, -0.864, -0.086, -0.421, -0.342, -0.248, -0.16, -0.089, -0.04, -0.021, -0.033, -0.021, -0.01, 0, 0.01, 0.02, 0.03, 0.04, 0.05, 0.061, 0.061, 0.071, 0.08, 0.1, 0.201, 0.21, 0.58, 0.874, 1.7, 2.6, and 3.4 kOe. (b) Voltage measured at 265 K from the data in (a) as a function of the magnetic field applied during MFC. A magnetic field as small as ~ 100 Oe is sufficient to switch the crystal from the "+1" single-domain state to the opposite "−1" single-domain state. A narrow intermediate field range produces multidomain configurations. (c) Temperature-dependent reverse magnetoelectric response measured after EA at 330 K with $E_A$ = -1.6 MV/m for 30 min, followed by MFC with different magnetic fields. MFC with -3.4 kOe prepares the initial "−1" single-domain state, followed by MFC sequences with -2.54, -1.664, -0.825, -0.61, -0.4, -0.198, -0.085, -0.61, -0.4, -0.198, -0.085, -0.009, 0.009, 0.119, 0.221, 0.418, 0.83, 1.73, 2.6, and 3.45 kOe. (d) Voltage measured at 265 K from the data in (c) as a function of the magnetic field applied during MFC. Small magnetic fields during MFC reversibly switch the crystal from the "-1" single-domain state to the "+1" single-domain state, demonstrating the influence of the residual internal electric bias on antiferromagnetic domain selection.

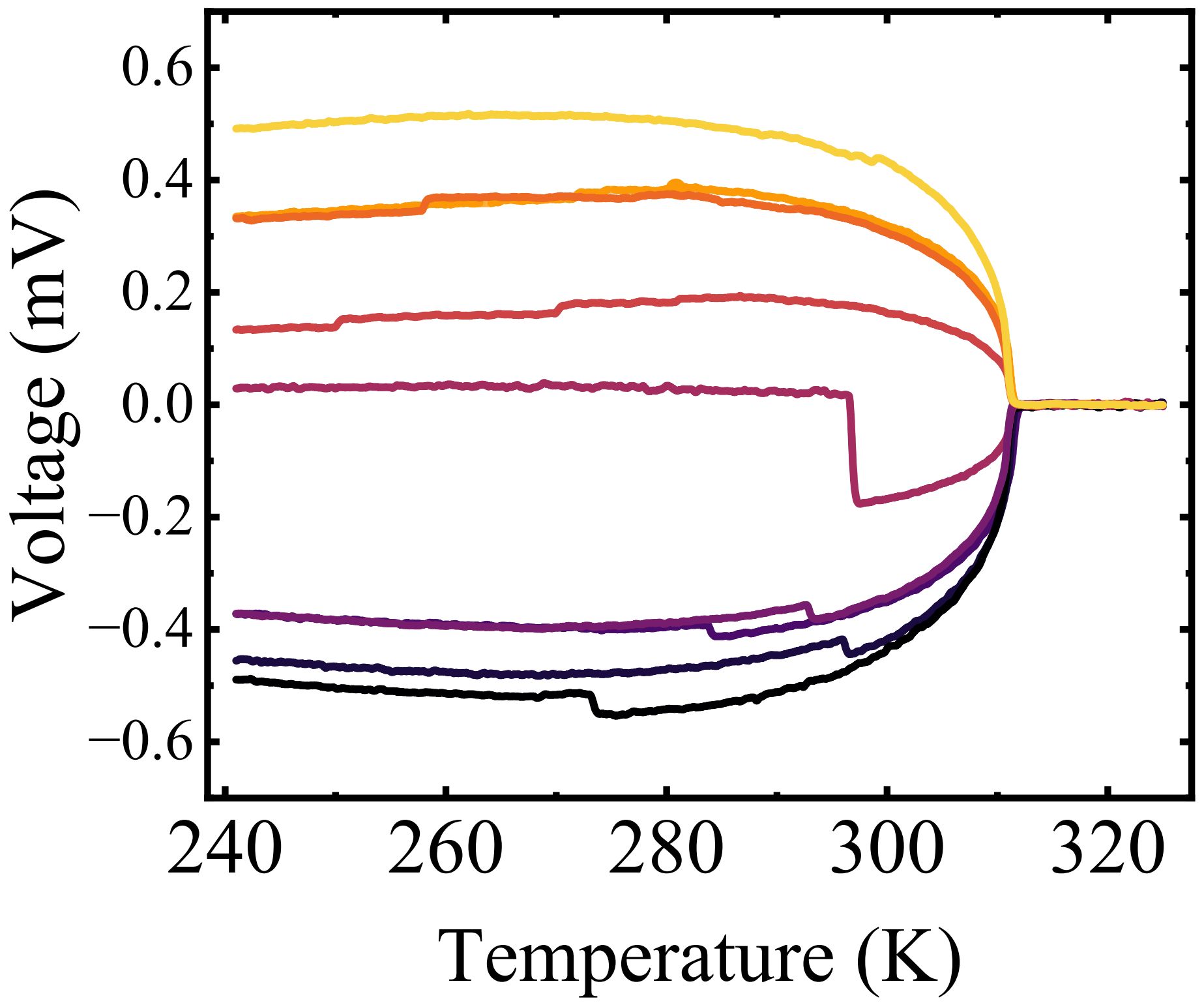

**Fig. S3. Spontaneous domain evolution below $T_N$.** Temperature-dependent reverse magnetoelectric response measured during warming from initially multidomain states. Abrupt changes in the signal are frequently observed, indicating spontaneous changes in the relative populations of antiferromagnetic domains during the temperature sweep. These discontinuous jumps are consistent with changes in domain populations, potentially arising from domain-wall motion and domain growth or shrinkage. In most cases, the signal changes in the direction of the majority domain, suggesting that the multidomain state progressively coarsens during warming.

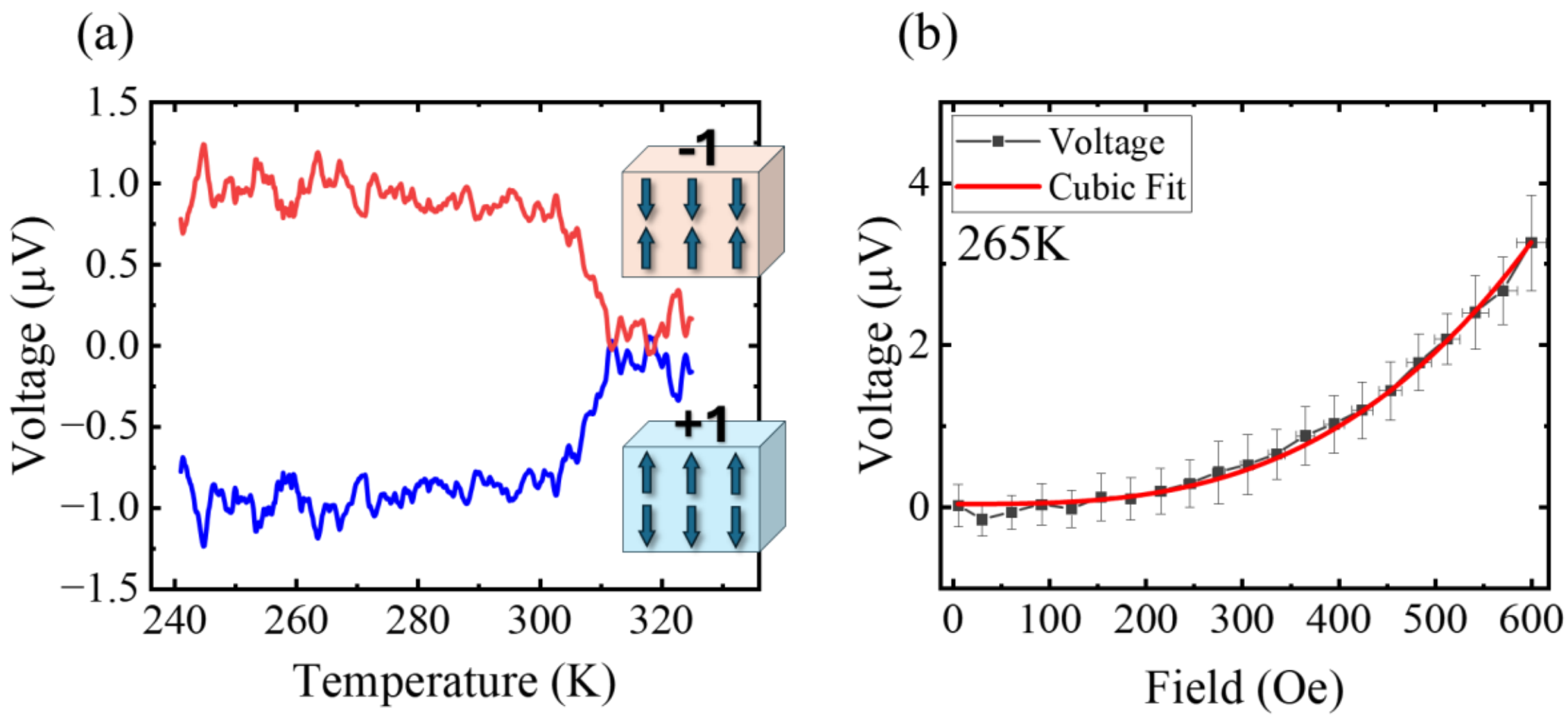

**Fig. S4. Higher-order magnetoelectric response.** Only a third-harmonic magnetoelectric signal was detected, which is likely associated with the fourth-order free-energy invariant proportional to $\boldsymbol{EH^3}$. Two representative measurements are shown. (a) Third-harmonic magnetoelectric response measured under identical conditions as the linear-response measurements. The signal amplitude is three orders of magnitude smaller than that of the first-order magnetoelectric response and exhibits the opposite sign of the corresponding magnetoelectric coefficient. (b) Third-harmonic voltage response as a function of the applied AC magnetic field amplitude, together with a cubic fit, $\boldsymbol{V_{(3\omega)} = 1.5 \times 10^{-14} H^3}$, confirming the expected nonlinear dependence on magnetic field.

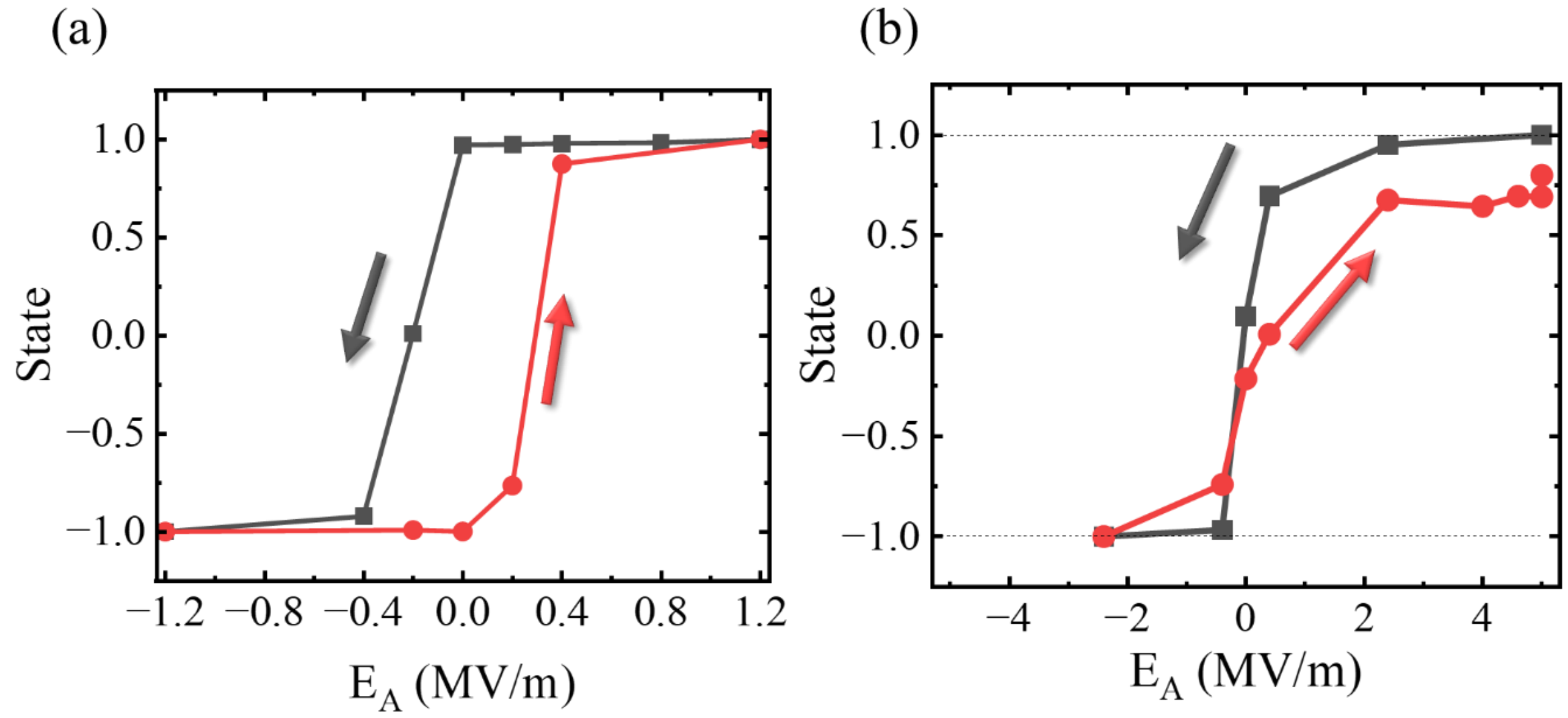

**Fig. S5. EA-field magnitude dependence for two $Cr_2O_3$ crystals.** All data presented in the main text and Supplemental Information, except those in (b), were obtained from crystal #1. Crystal #2 has dimensions of $5 \times 5 \times 0.25 mm^3$. In both (a) and (b), each data point was obtained after EA for 30 min at the indicated $E_A$, followed by MFC under $H = -3.4$ kOe. The electrodes were shorted after removal of $E_A$ and remained shorted during MFC. The vertical axis, labeled "State," represents the measured voltage normalized to the full response of the corresponding single-domain states, with $+1$ and $-1$ representing the two opposite Néel-vector orientations. (a) Crystal #1 exhibits a well-defined hysteresis loop with a coercive field of approximately 0.2 MV/m and saturation into both single-domain states. (b) Crystal #2 reproduces the EA-dependent switching behavior but exhibits a broader transition. A clear asymmetry is observed: positive $E_A$ requires a larger magnitude to approach saturation and achieve complete switching than negative $E_A$. The detailed origins of the differences between the two crystals are not yet fully understood. Possible contributing factors include differences in surface roughness, thermal history, and sample dimensions. In particular, crystal #1 was annealed in oxygen at 500 °C for 3 h, whereas crystal #2 was not subjected to this annealing treatment. These differences may influence the surface state and could therefore contribute to the distinct hysteresis characteristics observed in the two crystals. The remanence, hysteresis, and saturation observed in both crystals, despite quantitative differences in their field dependence, are consistent with a bistable polar response and place additional constraints on a simple transient-charge interpretation.